\documentclass[prl,twocolumn,floats,aps,epsfig,nofootinbib,amssymb]{revtex4}
\usepackage{subfigure}
\usepackage{graphicx}
\usepackage{cancel}
\usepackage{amssymb}
\usepackage{textcomp}
\usepackage{amsmath}
\usepackage{bm}
\usepackage{times}
\usepackage{epsfig}
\usepackage{color}
\def\beq{\begin{equation}}
\def\eeq{\end{equation}}
\def\bea{\begin{eqnarray}}
\def\eea{\end{eqnarray}}

	\newcommand{\abs}[1]{ \mathopen{}\left| {#1}\right| }
	\DeclareMathOperator{\real}{Re}
	\DeclareMathOperator{\imag}{Im}
	
\begin{document}

\title{\large MINIMAL GAUGED $U(1)_{B-L}$ MODEL WITH SPONTANEOUS R-PARITY VIOLATION}
\bigskip
\author{Vernon Barger}
\author{Pavel Fileviez P{\'e}rez}
\author{Sogee Spinner}
\address{
Department of Physics, University of Wisconsin, Madison, WI 53706, USA}
\date{\today}

\begin{abstract}
We study the minimal gauged $U(1)_{B-L}$ supersymmetric model and show
that it provides an attractive theory for spontaneous R-parity violation.
Both $U(1)_{B-L}$ and R-parity are broken by the vacuum expectation value
of the right-handed sneutrino (proportional to the soft SUSY masses),
thereby linking the $B-L$ and soft SUSY scales. In this context we find a
consistent mechanism for generating neutrino masses and a realistic mass spectrum,
all without extending the Higgs sector of the minimal supersymmetry
standard model. We discuss the most relevant collider signals and the connection
between the $Z^{'}$ gauge boson and R-parity violation.
\end{abstract}

\maketitle
\section{Introduction}
Supersymmetry (SUSY) is one of the most appealing solutions to the
hierarchy problem and candidates for physics beyond the Standard
Model (SM).  It's most minimal incarnation, the minimal supersymmetric
Standard Model (MSSM), contains two major puzzles which have
attracted the attention of many experts in the field. The first is the need
for a predictive mechanism for SUSY breaking and the second is the existence
of baryon and lepton number violating interactions. Pragmatically, the
so-called R-parity discrete symmetry is imposed in order to forbid these
interactions, which would lead to dimension four contributions to the decay
of the proton.  As a bonus, once this discrete symmetry is imposed, the lightest
supersymmetric particle (LSP) is stable and is a candidate for the
cold dark matter in the Universe. Therefore, it is important
to investigate the mechanisms for R-parity conservation.

The possible origin of such mechanisms in the MSSM have been studied in
detail in Ref.~\cite{Martin}.  Furthermore, extensions of the MSSM which
contain either a local or global $B-L$ symmetry will preserve R-parity at
low energies only if the fields which break that symmetry have even $B-L$
quantum numbers \cite{Martin}.  See Ref.~\cite{Goran} for the
study of this issue in $SO(10)$ grand unified theories and left-right models.

There are two simple frameworks where one can understand the origin
of the R-parity violating interactions: i) introducing new multiplets
having $U(1)_{B-L}$ as a global symmetry~\cite{GlobalBL} but one has
to face the Majoron problem~\cite{Majoron},
ii) the appealing possibility of having a local $U(1)_{B-L}$ symmetry where
the Majoron becomes the longitudinal component of the $B-L$ gauge boson and is
therefore no longer a problem.  Recently, this latter framework was used to construct
the simplest supersymmetric left-right theory~\cite{Spinner}. In this work we want to
study the predictions stemming from the simplest possible scenarios for spontaneous
R-parity breaking in light of this second framework.

To do this we follow a simple procedure: postulate the existence of an extra $U(1)_{B-L}$
local gauge symmetry and introduce only the fields necessary for a theoretically consistent
theory. These fields are the three generations of right-handed neutrinos required for
anomaly cancellation and we will introduce no other fields, \textit{i.e} the Higgs sector will be the same as
in the MSSM.  Once the right-handed sneutrino acquires a vacuum expectation value (VEV): $U(1)_{B-L}$
and R-parity are both broken reproducing the MSSM with baryon number
conservation but lepton number and R-parity violation and the $B-L$
scale is identified with the soft SUSY mass scale --- an interesting prediction.
Viable neutrino masses are generated through an extended seesaw mechanism
and a realistic spectrum is possible. This scenario leads to interesting phenomenology
due to the intimate relationship between the $Z^{'}$ gauge boson and R-parity
violation.  This is testable through the decays of the $Z^{'}$, and
the usual R-parity violating decays of neutralinos and charginos as well as the
SUSY induced mass degeneracy between the right-handed sneutrino and the $Z^{'}$ gauge boson.

This work is organized as follows: In Section I we present the model
and discuss the basic idea of R-parity violation. In Section II we study
the symmetry breaking mechanism showing the conditions on the soft SUSY
breaking terms. The full spectrum is discussed in Section III while
the possible smoking guns are presented in Section IV.
\section{$U(1)_{B-L}$ Extension of the MSSM}
It is well known that the MSSM allows for baryon and lepton number
violating interactions.  Usually, these terms are forbidden by imposing
the so-called R-parity. This discrete symmetry is defined as
$R=(-1)^{3(B-L)+2S}=(-1)^{2S} M$, where $M$ is the so-called Matter
parity. $M=-1$ for any matter superfield and $M=+1$ for the Higgs and
Gauge superfields. The MSSM R-parity violating interactions are
\begin{eqnarray}
\label{RPV}
{\cal W}_{RPV} &=& \epsilon_i \hat{L}_i \hat{H}_u \ + \ \lambda_{ijk} \hat{L}_i \hat{L}_j \hat{E}^C_k
\ + \ \lambda_{ijk}^{'} \ \hat{Q}_i \hat{L}_j \hat{D}^C_k \nonumber \\
& + & \lambda_{ijk}^{''} \ \hat{U}_i^C \hat{D}_j^C \hat{D}^C_k
\end{eqnarray}
where the first three terms violate the lepton number (L) and the last one violates the
baryon number (B). See Ref.~\cite{review} for the constraints coming from
proton decay and Refs.~\cite{Tao, Barbier} for previous studies.

However, forbidding all the terms in Eq.~(\ref{RPV}) is not necessary for a viable model.
In this work we want to understand the origin of the non-harmful R-parity violating interactions
in the simple extension of the MSSM by a local $U(1)_{B-L}$ symmetry. The full gauge group
is then $SU(3)_C \bigotimes SU(2)_L \bigotimes U(1)_Y \bigotimes U(1)_{B-L}$.
The matter chiral supermultiplets for quarks and leptons and their
($SU(2)_L, U(1)_Y, U(1)_{B-L}$) quantum numbers are given by
$
\hat{Q} = \left(
\begin{array} {c}
\hat{U} \\ \hat{D}
\end{array}
\right) \sim (2,1/3,1/3),
\ \ \
\hat{L} = \left(
\begin{array} {c}
 \hat{N} \\ \hat{E}
\end{array}
\right) \sim (2,-1,-1),
$
$
\hat{U}^C \sim (1,-4/3,-1/3),
\ \ \
\hat{D}^C \sim (1,2/3,-1/3),
\ \
\hat{E}^C \sim (1,2,1),
$
and in order to cancel the anomalies one introduces
three chiral superfields for the right-handed neutrinos:
$
\hat{N}^C \ \sim \ (1,0,1).
$
With this matter content, the superpotential reads as
\begin{equation}
{\cal W}_{BL}={\cal W}_{MSSM} \ + \ Y_\nu^D \ \hat{L}^T \ i \sigma_2 \ \hat{H}_u \ \hat{N}^C,
\end{equation}
where
\begin{eqnarray}
	{\cal W}_{MSSM} &=& Y_u \ \hat{Q}^T \ i \sigma_2 \ \hat{H}_u \ \hat{U}^C
\ + \ Y_d \ \hat{Q}^T \ i \sigma_2 \ \hat{H}_d \ \hat{D}^C \nonumber \\
& + & Y_e \ \hat{L}^T \ i \sigma_2 \ \hat{H}_d \ \hat{E}^C
\ + \ \mu \ \hat{H}_u^T \ i \sigma_2 \ \hat{H}_d.
\end{eqnarray}
The two MSSM doublet Higgses are defined by
\begin{equation}
\hat{H}_u = \left(
\begin{array} {c}
	\hat{H}_u^+
\\
	\hat{H}_u^0
\end{array}
\right) \ \sim \ (2, 1, 0),
\ \ \
\hat{H}_d = \left(
\begin{array} {c}
	\hat{H}_d^0
\\
	\hat{H}_d^-
\end{array}
\right) \ \sim \ (2, -1, 0).
\end{equation}
So far, it seems like extra superfields are needed to
break $U(1)_{B-L}$, since the Higgs doublets do not have $B-L$ quantum
numbers. As mentioned earlier, the choice for these fields depends on
whether or not the low energy theory should conserve R-parity (or M-Parity).
Therefore, R-parity conservation requires Higgs fields with an even value of $B-L$.
Typically, this is achieved by introducing several extra Higgs chiral superfields.

In this work, we wish to take advantage of the fact that the model already contains
scalar fields with the correct quantum numbers: the right-handed sneutrinos. Once one of
these fields acquires a VEV, it spontaneously breaks both the extra gauge symmetry,
$U(1)_{B-L}$, as well as R-parity and forces left-handed sneutrino, through mixing
terms, to acquire a VEV. Since lepton number is part of the gauge symmetry
the Majoron~\cite{Majoron} (the Goldstone boson associated with spontaneous
breaking of lepton number) becomes the longitudinal component of the $Z^{'}$
and does not pose a problem. Therefore, in this context one can have a simple and
consistent TeV scale theory for spontaneous $U(1)_{B-L}$
and R-parity violation with the same Higgs sector as the MSSM.

In addition to the superpotential, the model is also specified by the soft terms:
\begin{eqnarray}
	\nonumber
	V_{soft} & = & M_{\tilde N^C}^2 \abs{\tilde{N}^C}^2 \ + \ M_{\tilde L}^2 \ \abs{\tilde L}^2 + M_{\tilde E^C}^2 \ \abs{\tilde E^C}^2
	\\
	\nonumber
		& + & m_{H_U}^2 \abs{H_U}^2 + m_{H_D}^2 \abs{H_D}^2 \ + \ \left( \frac{1}{2} M_{BL} \tilde{B^{'}} \tilde{B^{'}} \right.
  	\nonumber
	\\
		& + &  \left. A_\nu^D \ \tilde{L}^T \ i \sigma_2 \ H_u \ \tilde{N}^C  \  + \  B\mu \ H_U^T \ i \sigma_2 \ H_D \right.
	\nonumber \\
		& + & \left. \mathrm{h.c.} \right) +...
\label{soft}
\end{eqnarray}
where the terms not shown here correspond to terms in the soft MSSM potential.
Now, we are ready to investigate the possible predictions coming from
spontaneous R-parity violation.
\section{Symmetry Breaking and R-parity Violation}
In this theory the gauge boson masses are generated by
the vacuum expectation values (VEVS) of sneutrinos
($\langle \tilde{\nu_i} \rangle=v_L^i/\sqrt{2}$
and $\langle \tilde\nu^C_i \rangle=v_R^i/\sqrt{2}$), and the Higgs doublets
($\left< H_u^0 \right> = v_u/\sqrt{2}$ and $\left< H_d^0 \right> = v_d/\sqrt{2}$).
The sneutrino VEVs also break R-parity and lepton number eliminating
the quantum numbers necessary to distinguish between the
lepton, Higgsino and gaugino sectors. Therefore the physical
charginos and neutralinos, as well as the Higgses will be admixtures
of these three sectors.

\textit{\underline{Symmetry Breaking}}:
The scalar potential in this theory is given by
\begin{eqnarray}
	V & = & V_{F} \ + \ V_D \ + \ V_{soft}^S,
\end{eqnarray}
where the relevant terms for $V_{soft}^S$ are given
in Eq.~(\ref{soft}). Once one generation of sneutrinos, $\tilde{\nu}$ and
$\tilde{\nu}^C$, $H_u$ and $H_d$, acquire a VEV, the scalar potential reads
\begin{eqnarray}
	\left<V_F \right> &=&
		\frac{1}{4} \left(Y_\nu^D \right)^2
		\left(
			v_R^2 v_u^2 + v_R^2 v_L^2 + v_L^2 v_u^2
		\right)
		+ \frac{1}{2} \mu^2
		\left(
			v_u^2 + v_d^2
		\right) \nonumber \\
		& -  & \frac{1}{\sqrt{2}} Y_\nu^D \ \mu \ v_R v_L v_d,
	\\
	\left<V_D \right> &=&
		\frac{1}{32}
		\left[
			g_2^2
			\left(
				 v_u^2 -v_d^2 - v_L^2
			\right)^2
			+ g_1^2
			\left(
				v_u^2 - v_d^2 - v_L^2
			\right)^2
		\right.
		\nonumber \\
		& + & \left.
			g_{BL}^2
			\left(
				v_R^2 - v_L^2
			\right)^2
		\right],
	\\
	\left<V_{soft}^S \right> &=&
		\frac{1}{2} M_{\tilde L}^2 v_L^2 + \frac{1}{2} M_{\tilde N^c}^2 v_R^2 + \frac{1}{2} M_{H_u}^2 v_u^2 + \frac{1}{2} M_{H_d}^2 v_d^2
		\nonumber \\
		& - & \text{Re} \left( B \mu \right) \ v_u v_d + \frac{1}{2 \sqrt{2}} \left(A_\nu^D + \left(A_\nu^D \right)^\dagger \right) \ v_R v_L v_u,
\nonumber
\\
\end{eqnarray}
where $g_1$, $g_2$ and $g_{BL}$ are the gauge couplings for $U(1)_Y$, $SU(2)_L$ and $U(1)_{B-L}$,
respectively. This can be minimized in the usual way but illuminating results can be found for
the case $v_R \gg v_u, v_d \gg v_L$, (a reasonable assumption given the
phenomenologically necessary hierarchy between the left-and right-handed scales):
\begin{equation}
	\label{MC.vR}
	v_R =
		\sqrt{\frac{- 8 M_{\tilde{N}^c}^2}{g_{BL}^2}},
	\ \ \
	v_L =
		\frac{B_\nu v_R}{M_{\tilde L}^2 - \frac{1}{8} g_{BL}^2 v_R^2},
\end{equation}
with $B_\nu \equiv \frac{1}{\sqrt{2}} \left(Y_\nu^D \ \mu \ v_d \ - \ A_\nu^D \ v_u \right)$.
The first part of Eq.~(\ref{MC.vR}) has the same form as the Standard Model minimization
condition and demonstrates the need for $M_{\tilde{N}^c}^2 < 0$, while the second part
indicates that $B_\nu$ should be small, \textit{i.e.}
$B_\nu \ll M_{\tilde L}^2 - \frac{1}{8} g_{BL}^2 v_R^2$ in order to have
$v_R \gg v_L$. The $v_u$ and $v_d$ minimization conditions are equivalent to
the MSSM ones in this limit.  

\textit{\underline{R-Parity Violation}}:
R-parity violating bilinear terms, which mix leptons with Higgsinos and gauginos,
will be generated from Yukawa couplings, after symmetry breaking and are given in
Table \ref{BRPV}.
\begin{table}[htdp]
\caption{Bilinear R-parity violating terms from: the gauge sector (left column) and the superpotential (right column).}
\begin{center}
\begin{tabular}{ccc}
\hline
\hline
$\frac{1}{2} g_{BL} \ v_R \ \left( \nu^c \ \tilde B^{'} \right)$ & $\quad\quad\quad$ &
$\frac{1}{\sqrt{2}} Y_\nu^D v_R \ \left( l^T i\sigma_2 \ \tilde{H}_u \right)$
\\
$\frac{1}{2} g_{2} \ v_L \ \left( \nu \ \tilde W^0 \right)$ & $\quad\quad\quad$ &
$\frac{1}{\sqrt{2}}Y_\nu^D \ v_L \ \left( \tilde{H}_u^0 \ \nu^c \right)$
\\
$\frac{1}{\sqrt{2}} g_{2} \ v_L \ \left( e \ \tilde W^+ \right)$
& $\quad\quad\quad$ & $\frac{1}{\sqrt{2}} Y_e \ v_L \ \left( \tilde{H}_d^- \ e^c \right)$
\\
$\frac{1}{2} g_{1} \ v_L \ \left( \nu \ \tilde B \right)$ & $\quad\quad\quad$ &
\\
\hline
\hline
\end{tabular}
\end{center}
\label{BRPV}
\end{table}%
The first term on the left is new and is the only term not suppressed by neutrino
masses. The first term on the right corresponds to the so-called $\epsilon$ term,
and the last three terms on the left are small but important for the decay
of neutralinos and charginos.

Trilinear R-parity violating interactions follow once the neutralinos are integrated out.
These will depend on the Yukawa couplings: $Y_e$ and $Y_d$.  The general
form for $\lambda$- and $\lambda^{'}$-type interactions is:
$Y_e /M_{\chi^0}$ and $Y_d /M_{\chi^0}$.
For realistic neutrino masses, these interactions are small enough to evade experimental bounds.
For further discussion on effective trilinear terms in bilinear R-parity violation see \cite{Barbier}.
\section{Mass Spectrum}
The purpose of this section is to check that a realistic spectrum exists.

{\textit{\underline{Gauge Bosons}}}:
The gauge sector consists of the SM gauge bosons and an extra neutral gauge boson,
the $Z^{'}$. In the case $v_R \gg v_u, v_d \gg v_L$ the extra gauge boson has the mass
$M_{Z^{'}}^2 = g_{BL}^2 v_R^2/4$. The most conservative bounds from LEP2
and CDF are $M_{Z^{'}}/g_{BL} > (5-10) $ TeV~\cite{PDG}. See
Ref~\cite{Petriello:2008zr} for a recent study of the $Z^{'}$ at
the LHC and Ref~\cite{Langacker:2008yv} for a review.
{\textit{\underline{Neutralinos and Neutrinos}}}:
Once R-parity is broken the neutralinos and neutrinos
$\left(\nu, \ \nu^c, \ \tilde B, \ \tilde B^{'},\ \tilde W_L^0, \ \tilde H_d^0, \ \tilde H_u^0 \right)$
mix. In order to understand the neutrino masses we
focus on the simple case $v_L \to 0$ and $Y_\nu^D$ small:
\begin{equation}
M_\nu = M_\nu^I + M_\nu^R,
\end{equation}
where $M_\nu^I$ is the type I seesaw contribution~\cite{TypeI}
and $M_\nu^R$  is due to R-parity violation. These contributions
are given by
\begin{eqnarray}
	M_\nu^I	&=&	\frac{1}{2} Y_\nu^D M_{\nu^C}^{-1} \left(Y_\nu^D\right)^T v_u^2,
	\\
	M_\nu^R	&=&	m \ M_{\tilde \chi^0}^{-1} \ m^T,
\end{eqnarray}
where
\begin{eqnarray}
M_{\nu^C} & \approx & \left(M_{BL} + \sqrt{4 M_{Z^{'}}^2 + M_{BL}^2}\right)/2 \\ 
m &= & \text{diag}\left(0,0,0,0,Y_\nu^D v_R/\sqrt{2}\right),
\end{eqnarray}
and $M_{\tilde \chi^0}$ is the neutralino mass matrix in the MSSM.
Therefore it is possible to simply reproduce
the light neutrino masses.

{\textit{\underline{Bosonic Spectrum}}}:

Defining the basis $\sqrt{2} \imag \left(\tilde \nu, \tilde \nu^c, H_d^0, H_u^0 \right)$
for CP-odd scalars, $\sqrt{2} \real \left(\tilde \nu, \tilde \nu^c, H_d^0, H_u^0 \right)$
for CP-even scalars and for the charged scalars $\left(\tilde e^*, \tilde e^c, H_d^{-*}, H_u^+ \right)$
we have computed the mass matrices. It is important to show that the spectrum of the theory is realistic and
the expected Goldstone bosons exist. Here, we analyze the spectrum
in the very illustrative limit of zero mixing between the left- and right-handed sneutrinos, \textit{i.e.}
$Y_\nu^D, A_\nu^D \rightarrow 0$. In this limit, Eq.~(\ref{MC.vR}) indicates
that $v_L \rightarrow 0$ as well and $B_\nu \rightarrow 0$, by definition.
Applying this limit to the bosonic mass matrices shows that they decouple into three
values: two eigenvalues representing the left- and right-handed slepton masses and the
MSSM two-by-two mass matrix for the up- and down-type Higgs. We will focus on the former
since the latter only reproduces the results of the MSSM.

One of the eigenvalues for the CP-odd sleptons corresponds to the Goldstone boson eaten by $Z^{'}$
(the Majoron) and is completely made up of the imaginary part of the right-handed sneutrino,
$\imag \tilde \nu^c$. The second eigenvalue is the mass of the physical complex
left-handed sneutrino (note that this is the physical state in this limit, as opposed to
the real and imaginary parts of the left-handed sneutrino):
\begin{equation}
	\label{mnuMass}
	m_{\tilde \nu}^2	=
		M_{\tilde L}^2 - \frac{1}{8} g_{BL}^2 v_R^2 - \frac{1}{8}\left(g_1^2 + g_2^2 \right) \left(v_u^2-v_d^2 \right).
\end{equation}
Note that this mass implies that $M_{\tilde L}^2 > \frac{1}{8} g_{BL} v_R^2$.
This mass is also an eigenvalue of the CP-even mass matrix.
The other CP-even eigenvalue is the mass for the
CP-even piece of the right-handed sneutrino, $\real \tilde \nu^c$:
\begin{equation}
	m_{\real \tilde \nu^c}^2	=
		g_{BL}^2 v_R^2/4,
\end{equation}
which is positive and is degenerate with the $Z^{'}$ mass.  Finally, the masses of the charged sleptons are:
\begin{align}
	\notag
	m_{\tilde e_L}^2 &	=
		M_{\tilde L}^2 - \frac{1}{8}g_{BL}^2 v_R^2 + \frac{1}{8} \left(g_2^2 - g_1^2 \right) \left(v_u^2 - v_d^2 \right) + \frac{1}{2} Y_e^2 v_d^2
\\
	m_{\tilde e_R}^2 &	=
		M_{\tilde E^c}^2 + \frac{1}{8}g_{BL}^2 v_R^2 + \frac{1}{4} g_1^2 \left(v_u^2 - v_d^2 \right) + \frac{1}{2} Y_e^2 v_d^2.
\end{align}
A closer examination of Eqs.(16,18) for the MSSM fields
indicates that these values are the MSSM mass values modified appropriately by $B-L$
$D$-term contributions. All of these masses are realistic given $M_{\tilde L}^2 > \frac{1}{8} g_{BL} v_R^2$.
Of further interest is the prediction of the degeneracy between the $Z^{'}$ gauge boson and the
physical right-handed sneutrino.  This degeneracy also extends to the right-handed neutrino when
$g_{BL} v_R \gg M_{BL}$.  Corrections to the approximate masses presented here would be of
the order $Y_\nu^D v_L v_R$ or $A_\nu^D v_L \frac{v_R}{v_u}$.
All of these terms are highly suppressed due to neutrino masses, making this discussion
relevant even in the non-limit case.

{\textit{\underline{Charginos and Charged Leptons}}}:
Mixing between the charged leptons and the charginos will occur
in the charged fermion sector, $\left(e^c, \ \tilde W_L^+, \ \tilde H_u^+ \right)$ and
$\left( \ e, \ \tilde W_L^-, \ \tilde H_d^-\right)$. Since the mixing between the MSSM charginos and the
charged leptons is proportional to $v_L$ and $Y_\nu^D$ small corrections to the charged
lepton masses can exist once the charginos are integrated out.
However, this contribution is always small once we impose the neutrino constraints.
\section{Collider Signals}
As a consequence of R-parity violation, the lightest neutralino will be
unstable and will decay via lepton number violating interactions. These type
of interactions will also exist for the charginos and the new gauge boson:

\textit{Sleptons decays}: It is important to emphasize the importance
of the lepton number violating decays of sleptons. Here one has the decays
$\tilde{\nu} \to \nu \nu, e^+_i e^-_j$, $\tilde{e}_i \to e_j \nu_k$,
$\tilde{e}^c_i \to e_j^+ \bar{\nu}^c_k$ and $\tilde{\nu^c} \to e \tilde{e}^c$. These decays
are proportional to $v_L$ or $Y_\nu^D v_R$ and are crucial for the test of the model.

\textit{$Z^{'}$ decays}: The $Z^{'}$ will decay mainly into (s)leptons since its coupling to
(s)quarks is suppressed by the corresponding $B-L=1/3$. In addition to the typical $Z^{'}$ decays,
new lepton number violating decays will be possible. These include
$Z^{'} \rightarrow e_j^\pm \tilde \chi_j^\mp$ which are suppressed by $v_L$.
Also possible are the very interesting decays $Z^{'} \to \overline{\nu^C} \nu^C$,
where the right-handed neutrinos can decay mainly to an electron and a selectron.
These decays are lepton number violating and proportional to $v_R$.
In particular very exciting signals are possible if the selectron is long-lived.

\textit{Neutralino decays}:
These will include $\tilde \chi_i^0 \rightarrow Z^0 \bar \nu$ and
$\tilde \chi_i^0 \rightarrow W^\pm e^\mp$ as usual. In the case when
the neutralino is the up-like Higgsino, these decays are proportional
to $v_R$, while in the rest of the cases are suppressed by $v_L$.

\textit{L-violating Higgs decays}:
The Higgses now could have lepton number violating
decay channels open such as MSSM-like Higgs
into a slepton and a $W$ or $Z$ if kinematically allowed.

\textit{Chargino decays}:
In this case  new decays into charged leptons and a $Z$ or $W$ exist.
All these decays are suppressed by $v_L$ or $Y_\nu^D$ once we impose
the constraints coming from neutrino masses. The properties of all
decays mentioned above will be studied
in a future publication. See Ref.~\cite{Marco} for a
recent study of R-parity violating decays at the LHC.
\section{Summary and Outlook}
We proposed and studied a mechanism for spontaneous
R-parity violation in the minimal $U(1)_{B-L}$ extension
of the MSSM. The symmetry breaking is achieved through the
VEV of the right-handed sneutrino so no new Higgs fields are
needed. Because the sneutrino VEV is proportional to the soft
SUSY mass scale, the $B-L$ breaking scale will be connected
to the soft SUSY mass scale. The generation of viable neutrino
masses, properties of the spectrum, the generation of R-parity
breaking terms and the possible signals at future colliders
experiments have been discussed and the
mass degeneracy between the $Z^{'}$ and the right-
handed sneutrino has been noted. The fact that such a rich model
is a product of simply extending the MSSM by local $U(1)_{B-L}$
and provides an understanding of spontaneous breaking of R-parity
is very appealing to us and we plan to study further it's
phenomenology in a future publication.

{\textit{Acknowledgments}}:
{\small P.F.P. and S.S. would like to thank Espresso Royale for hospitality.
We would also like to thank Manuel Drees, Tao Han and Frank Petriello
for useful discussions. The work of P.F.P. was supported in part by
the U.S. Department of Energy contract No. DE-FG02-08ER41531 and in part
by the Wisconsin Alumni Research Foundation. V.B. and S.S. are supported in
part by the U.S. Department of Energy under grant No. DE-FG02-95ER40896,
and by the Wisconsin Alumni Research Foundation.}



\begin{thebibliography}{000}

\bibitem{Martin}
  See for example:
 R.~N.~Mohapatra,
  Phys.\ Rev.\  D {\bf 34}, 3457 (1986);
  S.~P.~Martin,
  Phys.\ Rev.\  D {\bf 46} (1992) 2769;
  Phys.\ Rev.\  D {\bf 54} (1996) 2340.

\bibitem{Goran}
  C.~S.~Aulakh, B.~Bajc, A.~Melfo, A.~Rasin and G.~Senjanovic,
  Nucl.\ Phys.\  B {\bf 597} (2001) 89;
K.~S.~Babu and R.~N.~Mohapatra,
  Phys.\ Lett.\  B {\bf 668} (2008) 404.


\bibitem{GlobalBL}
  A.~Masiero and J.~W.~F.~Valle,
  Phys.\ Lett.\  B {\bf 251} (1990) 273;
  J.~C.~Romao, C.~A.~Santos and J.~W.~F.~Valle,
  Phys.\ Lett.\  B {\bf 288} (1992) 311;
  M.~Shiraishi, I.~Umemura and K.~Yamamoto,
  Phys.\ Lett.\  B {\bf 313} (1993) 89;
  M.~Chaichian and R.~Gonzalez Felipe,
  Phys.\ Rev.\  D {\bf 47} (1993) 4723;
  M.~Chaichian and A.~V.~Smilga,
  Phys.\ Rev.\ Lett.\  {\bf 68} (1992) 1455.

\bibitem{Majoron}
  G.~B.~Gelmini and M.~Roncadelli,
  Phys.\ Lett.\  B {\bf 99} (1981) 411;
  C.~S.~Aulakh and R.~N.~Mohapatra,
  Phys.\ Lett.\  B {\bf 119} (1982) 136;
  G.~G.~Ross and J.~W.~F.~Valle,
  Phys.\ Lett.\  B {\bf 151} (1985) 375;
  M.~C.~Gonzalez-Garcia and J.~W.~F.~Valle,
  an additional Z,''
  Nucl.\ Phys.\  B {\bf 355} (1991) 330;
  Y.~Chikashige, R.~N.~Mohapatra and R.~D.~Peccei,
  Phys.\ Lett.\  B {\bf 98} (1981) 265;
A.~Santamaria and J.~W.~F.~Valle,
  Phys.\ Lett.\  B {\bf 195}, 423 (1987).

\bibitem{Spinner}
  P.~Fileviez P\'erez and S.~Spinner,
  Phys.\ Lett.\  B {\bf 673} (2009) 251.

\bibitem{review}
  P.~Nath and P.~Fileviez P\'erez,
  Phys.\ Rept.\  {\bf 441} (2007) 191;
  P.~Fileviez P\'erez,
  J.\ Phys.\ G {\bf 31} (2005) 1025.

\bibitem{Tao}
  V.~D.~Barger, G.~F.~Giudice and T.~Han,
  Phys.\ Rev.\  D {\bf 40} (1989) 2987.
\bibitem{Barbier}
  R.~Barbier {\it et al.},
  Phys.\ Rept.\  {\bf 420} (2005) 1.


\bibitem{PDG}
  C.~Amsler {\it et al.}  [Particle Data Group],
  Phys.\ Lett.\  B {\bf 667} (2008) 1.

\bibitem{Petriello:2008zr}
  F.~Petriello and S.~Quackenbush,
  Phys.\ Rev.\  D {\bf 77}, 115004 (2008).

  \bibitem{Langacker:2008yv}
  P.~Langacker,
  arXiv:0801.1345 [hep-ph].

  \bibitem{TypeI}
  P.~Minkowski,
  Phys.\ Lett.\ B {\bf 67} (1977) 421;
  T. Yanagida,
p.~95, KEK Report 79-18, Tsukuba (1979);
  M. Gell-Mann, P. Ramond and R. Slansky,
   in {\it Supergravity}, eds. P. van Nieuwenhuizen et al.,
   (North-Holland, 1979), p.~315;
  S.L. Glashow, in {\it Quarks and Leptons}, Carg\`ese, eds. M. L\'evy et al.,
(Plenum, 1980), p. 707;
  R.~N.~Mohapatra and G.~Senjanovi\'c,
  Phys.\ Rev.\ Lett.\  {\bf 44} (1980) 912.

  \bibitem{Marco}
  F.~de Campos, M.~A.~Diaz, O.~J.~P.~Eboli, M.~B.~Magro, W.~Porod and S.~Skadhauge,
  Phys.\ Rev.\  D {\bf 77} (2008) 115025.

\end{thebibliography}
\end{document}